\documentclass{fmj2010}
\usepackage{graphicx}

\def\fermilat{\textit{Fermi}/LAT}
\def\fermi{\textit{Fermi}}
\def\agile{\textit{AGILE}}
\begin{document}
  \title{Monitoring of $\gamma$-ray blazars with AGILE}

   \author{F.~D'Ammando\inst{1} on behalf of the \agile\, Team} 

   \institute{$^1$ INAF--IASF Palermo, Via Ugo La Malfa 153, I-90146 Palermo,
   Italy}

   \abstract{Thanks to the wide field of view of its $\gamma$-ray imager, the
   \agile\, satellite obtained a long term monitoring of the brightest blazars
   in the sky and during the first 3 years of operation detected several
   blazars in a high $\gamma$-ray state: 3C 279, 3C 454.3, PKS 1510$-$089, S5
   0716$+$714, 3C 273, W Comae, and Mrk 421. Through the rapid
   dissemination of our alerts we were able to obtain also multi-wavelength
   data from many observatories such as {\textit{Spitzer}}, {\textit{Swift}},
   RXTE, {\textit{Suzaku}},
   XMM-{\textit{Newton}}, INTEGRAL, MAGIC, VERITAS, and ARGO as well as radio-to-optical
   coverage by means of the MOJAVE project, the GASP project of the WEBT and
   the REM Telescope. This large coverage over the whole electromagnetic
   spectrum gave us the opportunity to study the variability correlations
   between the emission at different frequencies and to build truly
   simultaneous spectral energy distributions of these sources from radio to
   $\gamma$-rays, investigating in detail the emission mechanisms of blazars and
   uncovering in some cases a more complex behaviour with respect to the
   standard models. We present an overview of the most interesting \agile\,results on these $\gamma$-ray blazars and the relative multiwavelength data.}

   \maketitle
%

\section{Introduction}

Blazars constitute the most enigmatic subclass of Active Galactic Nuclei
(AGNs), characterized by the
emission of strong non-thermal radiation across the entire electromagnetic spectrum and in particular intense and variable
$\gamma$-ray emission above 100 MeV (\cite{Hart99}). The typical
observational properties of blazars include
irregular, rapid and often very large variability, apparent super-luminal
motion, flat radio spectrum, high and variable polarization at radio and
optical frequencies. These features are interpreted as the result of the
emission of electromagnetic radiation from a relativistic jet that is viewed
closely aligned to the line of sight (\cite{BR}; \cite{UP}). 

Blazars emit across several decades of energy, from radio
to TeV energy bands, and thus they are the perfect candidates for simultaneous
observations at different wavelengths. Multi-wavelength studies of variable
$\gamma$-ray blazars have been carried out 
since the beginning of the 1990s, thanks to the EGRET instrument onboard $Compton$ $Gamma$-$Ray$ $Observatory$
$(CGRO)$, providing the first evidence that the Spectral Energy Distributions (SEDs) of
the blazars are typically double humped with the first peak occurring in the
IR/optical band in the so-called $red$ $blazars$ (including Flat Spectrum Radio
Quasars, FSRQs, and Low-energy peaked BL Lacs, LBLs) and in UV/X-rays in the
so-called $blue$ $blazars$ (including High-energy peaked BL Lacs, HBLs). 

The first peak is interpreted as synchrotron radiation from high-energy electrons in a
relativistic jet. The SED second component, peaking at MeV--GeV energies in
$red$ $blazars$ and at TeV energies in $blue$ $blazars$, is commonly
interpreted as inverse Compton (IC) scattering of seed photons, internal or external to the jet, by highly relativistic
electrons (\cite{Ul}), although other models involving hadronic
processes have been proposed (see e.g.~\cite{Bo} for a recent review). 

With the detection of several blazars in $\gamma$-rays by EGRET (\cite{Hart99})
the study of this class of object has made significant progress. In fact,
considering that the large fraction of the total power of blazars is emitted
in the $\gamma$-rays, information in this energy band is crucial to study the
different radiation models. 3C 279 is the best example of multi-epoch studies at different frequencies
performed by EGRET during the period 1991--2000 (\cite{Har01}). Nevertheless,
only a few objects were detected in the EGRET {\textit{era}} on a time scale
of few weeks in $\gamma$-rays and simultaneously monitored at different
energies to obtain a wide multi-frequency coverage.

The interest in blazars is now even more renewed
thanks to the simultaneous presence of two $\gamma$-ray satellites, \agile\, and
\fermi\,, and the possibility to obtain $\gamma$-ray observations
over long timescales simultaneously with data collected from radio to TeV
energies allowing us to reach a deeper insight on the jet structure and the
emission mechanisms at work in blazars. 

\begin{figure*}[hhht!]
\sidecaption
\includegraphics[width=13.4cm]{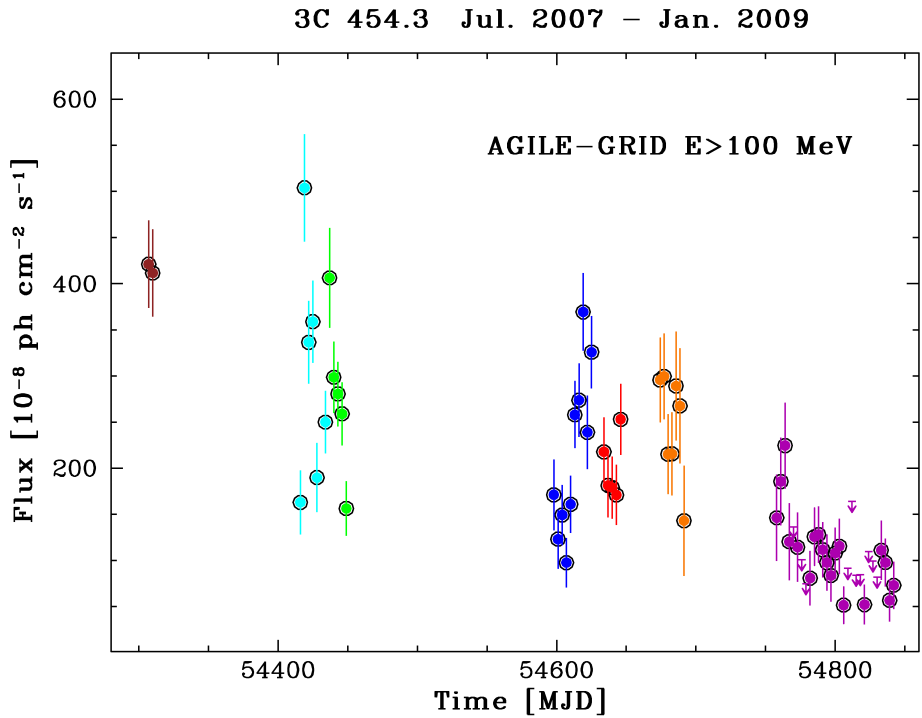}
  \caption{ \,\agile\, GRID light curve of 3C 454.3 collected between July 2007 and
  January 2009 at $\sim$3 day resolution in
  units of 10$^{-8}$ ph cm$^{-2}$ s$^{-1}$. [Adapted from \cite{V18}].} 
 \end{figure*}

\section{Blazars and AGILE}

The $\gamma$-ray observations of blazars are a key scientific project of
the \agile\, ({\it Astrorivelatore Gamma ad Immagini LEggero}) satellite (\cite{Tav09}). Thanks to the wide field of view of its
$\gamma$-ray imager ($\sim$ 2.5
sr), \agile\ monitored tens of potentially $\gamma$-ray emitting AGNs during each
pointing and after November 2009 in the new spinning mode the number of
$\gamma$-ray sources simultaneously monitored by the satellite is still increased. 

In the first 3 years of operation, \agile\ detected several blazars during
high $\gamma$-ray activity and extensive multi-wavelength campaign were
organized for many of them, providing the possibility to monitoring on long
timescales the brightest objects. The $\gamma$-ray activity timescales of these blazars goes from a few days (e.g.,~S5
0716$+$714 and 3C 273) to several weeks (e.g.,~3C 454.3 and PKS 1510$-$089) and the flux
variability observed has been negligible (e.g.,~3C 279), very rapid (e.g.,~PKS
1510$-$089) or extremely high (e.g.,~3C 454.3 and PKS 1510$-$089). However, we note that only a few objects were detected more than once in flaring state by \agile\ and mostly already known $\gamma$-ray emitting
sources showed intense flaring activity. This evidence, together with the results
of \fermilat\, obtained during the first 11 months of operation (\cite{Abdo2}), suggest possible constraint on the
properties of the most intense $\gamma$-ray emitters. In the following
we will present the most interesting results on the studies of the individual
sources detected by \agile\,.

\section{Individual Sources}

\subsection{3C 454.3}

Among the FSRQs 3C 454.3 is one of the brightest object and also the source that exhibited
the most variable activity in the last years. In particular during May 2005, 3C 454.3 was reported to undergo a very strong
optical flare (Villata et al.~2006). This exceptionally high state triggered
observations by high-energy satellites (RXTE: \cite{remillard}; {\it Chandra}:
\cite{vil06}; INTEGRAL: \cite{Pian2006:3C454_Integral}; {\it Swift}: \cite{Giommi2006:3C454_Swift}) which confirm an
exceptionally high flux also in X-ray band. Unfortunately, no $\gamma$-ray
satellite was operative at that time. 

In mid-July 2007, 3C~454.3 underwent a new optical brightening that triggered
observations at all frequencies, including a Target of Opportunity (ToO) by
the \agile\, $\gamma$-ray satellite (\cite{Vercellone2008}). That was the
beginning of an extraordinary long-term $\gamma$-ray activity of the source until the huge $\gamma$-ray flare observed in early
December 2009 (Striani et al.~2010). 
In the period July 2007--January 2009 the \agile\ satellite monitored
intensively 3C 454.3 together with {\it Spitzer}, GASP-WEBT, REM, MITSuME,
{\it Swift}, RXTE, {\it Suzaku} and INTEGRAL observatories, with two dedicated
campaigns organized during November 2007 and December 2007, as reported
respectively in \cite{Ver09} and \cite{Don09}, and yielding the longest multi-wavelength coverage of this $\gamma$-ray quasar so
far (\cite{V18}). 

\begin{figure}
\centering
\includegraphics[width=8.75cm]{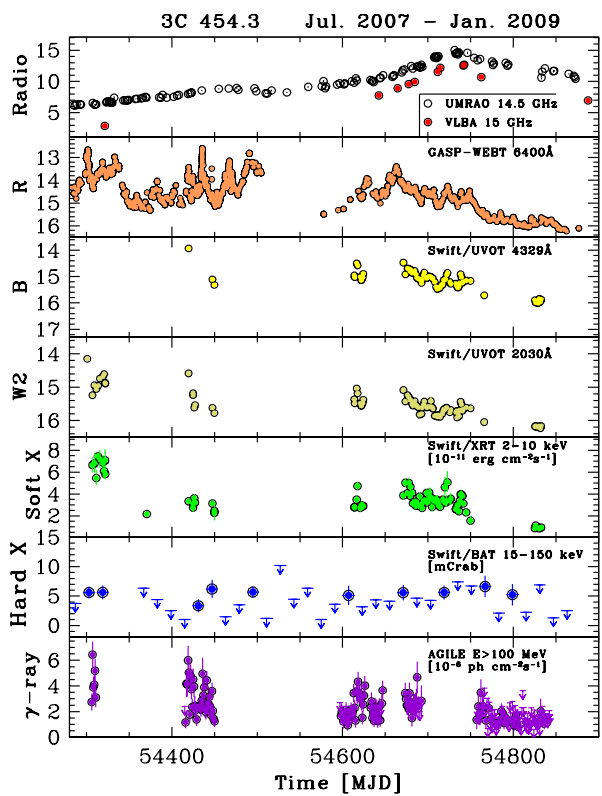}
  \caption{3C 454.3 light curves between July 2007 and January 2009 at
  increasing energies from top to bottom. Data were collected by \agile\,, {\textit{Swift}} (BAT, XRT and UVOT), GASP-WEBT, and VLBA. [Adapted from \cite{V18}].} 
 \end{figure}

The source underwent an unprecedented long period of high $\gamma$-ray
activity, showing flux levels variable on short timescales of 24--48 hours and
reaching on daily timescale a $\gamma$-ray flux higher than 500$\times$10$^{-8}$ ph cm$^{-2}$ s$^{-1}$ (Fig.~1 and
Fig.~2, bottom panel). A dimishing trend of the $\gamma$-ray flux from July
2007 and January 2009 was observed with a hint of
``harder-when-brighter'' behaviour,  previoulsy observed in $\gamma$-rays only
for 3C 279 in the EGRET {\textit{era}} (\cite{Har01}). Also the optical
flux appears extremely variable with a brightening of several tenths of
magnitude in a few hours. Emission in optical range appears to be correlated
with that at $\gamma$-rays, with a lag of the $\gamma$-ray flux with respect
to the optical one less than 1 day during bright states. However,
overimposed to the overall trend some sub-structures on shorter timescales
with different variability could be present in optical and $\gamma$-rays.

From the comparison of the light curves from radio to $\gamma$-rays shown
in Fig. 2 it is noticeable that, while at almost all the
frequencies the flux shows a diminishing trend with time during the
period July 2007--January 2009, the 15 GHz radio core flux increases, although
no new jet component seems to be detected in the high resolution VLBA
images. The different behaviour observed in radio, optical and $\gamma$-rays
from the end of 2007 could be interpreted in the framework of a helical jet
model as a change in the jet geometry between 2007 and 2008.  

The dominant emission mechanism above 100 MeV in 3C 454.3 seems to be the IC scattering of relativistic electrons in the jet on the
external photons from the Broad Line Region (BLR), even if in some cases also
the contribution of external Compton (EC) of seed photons from a hot corona
could be not negligible (Donnarumma et al. 2009b). 

\begin{figure*}[hhht!]
\sidecaption
\includegraphics[width=13.15cm]{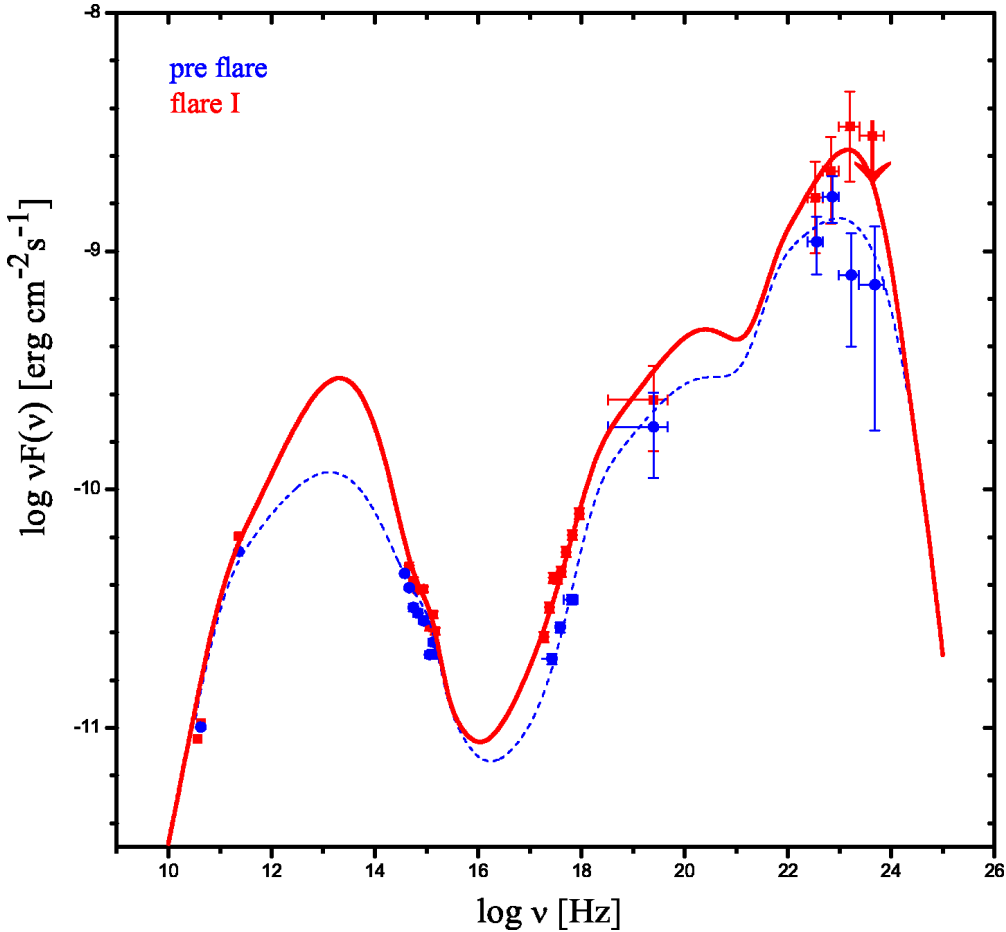}
  \caption{\,Spectral energy distribution of 3C 454.3 in the pre-flare and
  super-flare periods during December 2009. In the modeling of the SED
  relative to the super-flare an additional component is
  added. [Adapted from \cite{Pac10}].} 
 \end{figure*}

During December 2009, 3C 454.3 became the brightest $\gamma$-ray
source in the sky, reaching a peak flux of about 2000 $\times$ 10$^{-8}$ ph
cm$^{-2}$ s$^{-1}$ on 2--3 December 2009. Intensive multi-frequency
observations showed an overall correlation at all wavelengths for both long
and short timescales. However, the unusual $\gamma$-ray super-flaring activity
is not accompanied by strong emission of similar intensity in the optical or
even in the soft X-ray bands. The pre- and post-flare broad band behaviour can
be adequately represented by a simple one-zone synchrotron self Compton (SSC) model plus EC in which the accretion disk and the BLR provide the necessary
soft radiation field for the IC components. Instead, the spectrum of the
2-3 December 2009 super-flare would require with respect to the pre-flare an increase of the electron energy
and density and a slight reduction of the comoving magnetic field for the whole
electron population of the blob (see also Bonnoli et al. 2010). We use a
different approach, assuming a long-term rise and fall of the accretion rate onto the
central black hole that causes an overall increase of the synchrotron emission
and of the soft photon background scattered. An additional population of
electrons, due to an additional particle acceleration and/or plasmoid ejection
near the jet basis, could be present during the super-flare (Fig. 3; \cite{Pac10}).

\subsection{PKS 1510--089}

PKS 1510$-$089 is another blazar that in the last three years showed high
variability over all the electromagnetic spectrum, in particular high
$\gamma$-ray activity was observed by \agile\ and \fermi. \agile\
detected intense flaring episodes in August 2007 (\cite{Puc08}) and
March 2008 (\cite{DAm09a}) and an extraordinary
actitivity during March 2009 (\cite{DAm09b}). 

During the period 1--16 March 2008, \agile\ detected
an average flux from PKS 1510$-$089 of (84 $\pm$ 17)$\times$10$^{-8}$ ph cm$^{-2}$ s$^{-1}$ for E $>$ 100 MeV. The flux measured between 17 and 21 March
was a factor of 2 higher, with a peak level of (281 $\pm$ 68)$\times$10$^{-8}$ ph cm$^{-2}$ s$^{-1}$ on 19 March 2008. Moreover, between
January and April 2008 the source showed an intense and variable optical
activity with several flaring episodes of fast variability. A significant
increase of the flux was observed also at submillimetric frequencies in mid April, suggesting that the mechanisms producing the flaring events in
the optical and $\gamma$-ray bands also interested the submillimetric zone, with
a delay. 

The $\gamma$-ray flare triggered 3 {\textit{Swift}} ToO observations in
three consecutive days between 20 and 22 March 2008. The first XRT observation
showed a very hard X-ray photon index ($\Gamma$ = 1.16 $\pm$ 0.16) with a flux
in the 0.3--10 keV band of (1.22 $\pm$ 0.17)$\times$10$^{-11}$ erg cm$^{-2}$ s$^{-1}$ and a decrease of the flux of about 30$\%$ between 20 and
21 March. The $Swift$/XRT observations show a harder-when-brighter
behaviour of the spectrum in the X-ray band, confirming a behaviour already
observed in this source by \cite{Ka}, a trend usually
observed in HBL but quite rare in FSRQs such as PKS 1510$-$089. 

\noindent This harder-when-brighter behaviour is likely due to the different
variability of the SSC and EC components, therefore to the change of the
relative contribution of each component. Thus, the X-ray photon index observed
on 20 March could be due to the combination of SSC and EC emission and
therefore to the mismatch of the spectral slopes of these two components. The SED for the \agile\ observation of 17--21 March 2008 together with the
simultaneous data collected from radio-to-X-rays by
GASP-WEBT and {\textit{Swift}} is modelled with thermal emission
of the disc, SSC model plus the contribution by EC scattering of direct disc
radiation and of photons reprocessed by the BLR (see Fig.~4). Some features in
the optical/UV spectrum indicate the presence of Seyfert-like components, such as the little and big blue bumps.

PKS 1510$-$089 showed an extraordinary $\gamma$-ray activity during March
2009, with several flaring episodes and a flux that reached 600$\times$10$^{-8}$ ph cm$^{-2}$ s$^{-1}$ (Fig.~5). During February--March 2009 the source also showed an increasing activity in optical
and near-IR, with a flaring episode on 26--27 March 2009 observed by GASP-WEBT
and REM. Instead the
{\textit{Swift}}/XRT observations show no clear correlation of the X-ray
emission with the optical and $\gamma$-ray ones. In Fig. 6 we compare the SED
from radio-to-UV for 25--26 March 2009 with those collected on 20--22 March 2008 and 18 March 2009. The SED collected on 18
March 2009 confirmed the evidence of thermal signatures in the optical/UV
spectrum of PKS 1510--089 also during high $\gamma$-ray states. On the other hand, taking into
account that the dip at W1 could be systematic, the broad band spectrum from
radio-to-UV during 25--26 March 2009 show a flat spectrum in the optical/UV
energy band, suggesting an important contribution of the synchrotron emission
in this part of the spectrum during the larger $\gamma$-ray flare and
therefore a significative shift of the synchrotron peak, usually observed in
this source in the infrared. The increase of the synchrotron emission leads to
the decrease of the evidence of the thermal features observed in the other SEDs.

\begin{figure*}[!ht]
\sidecaption
\includegraphics[width=12.275cm]{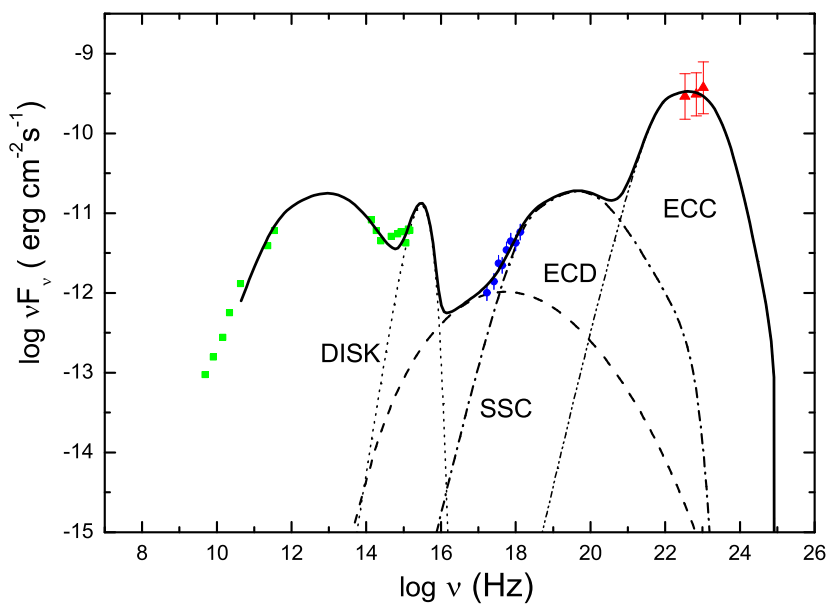}
  \caption{\,SED of PKS 1510$-$089 on mid-March 2008 with \agile\,, {\textit{Swift}} and GASP-WEBT data. The dotted, dashed,
  dot-dashed, and double dot-dashed lines represent the accretion disk emission, the SSC, the external Compton on the disk radiation (ECD) and on the BLR radiation (ECC), respectively. [Adapted from \cite{DAm09a}].} 
 \end{figure*}

\begin{figure*}[!ht]
\sidecaption
\includegraphics[width=12.275cm]{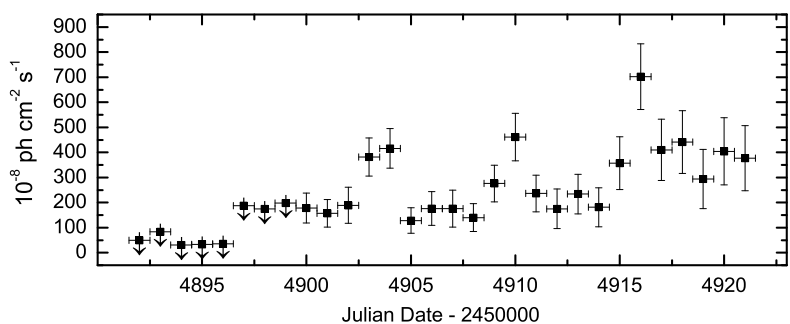}
  \caption{\,\agile\ light curve of PKS 1510$-$089 between 1 and 30
  March 2009 for E $>$ 100 MeV. The downward arrows
  represent 2-$\sigma$ upper limits. [Adapted from \cite{DAm09b}].} 
 \end{figure*}

\subsection{S5 0716+714}

The intermediate BL Lac (IBL) object S5 0716+714 was observed by \agile\,
during two different periods: 4--23 September and 23 October--1 November 2007, as
discussed in \cite{Chen}. In particular, between 7 and 12 September 2007 the source showed a high $\gamma$-ray activity with an average
flux of F$_{\rm E > 100\, MeV}$ = (97 $\pm$ 15)$\times$10$^{-8}$ ph cm$^{-2}$ s$^{-1}$
and a peak level of F$_{\rm E>100\, MeV}$ =
(193 $\pm$ 42)$\times$10$^{-8}$ ph cm$^{-2}$ s$^{-1}$, with an increase of flux by
a factor of four in three days. 
The flux detected by \agile\,
is the highest ever detected from this object and one of the most high flux observed from
a BL Lac object. A simultaneous GASP-WEBT optical campaign was performed and the resulting SED is well consistent with a
two-components SSC model (see Fig. 7). Another very intese $\gamma$-ray flare,
with a flux of the order of 200$\times$10$^{-8}$ ph cm$^{-2}$ s$^{-1}$, was
detected by \agile\ on 22-23 September 2007.

Recently, \cite{Ni} estimated the redshift of the source (z = 0.31 $\pm$ 0.08) and this allowed us to calculate the total power transported in the jet, which
results extremely high for the two flaring episodes, approaching or slightly
exceeding the maximum power generated by a spinning BH of 10$^{9}$ M$_\odot$
through the pure Blandford-Znajek mechanism (\cite{Vit09}). If confirmed, this
violation could be explained in terms of the alternative Blandford-Payne
mechanism (\cite{BP}) that, however, requires an ongoing accretion not
supported by the observations of S5 0716$+$714. Alternatively, a so high power
could be due to a less conservative value of the magnetic field related to
particle orbits plunging from the disk toward the BH horizon
(\cite{meier2002}) into a region with strong gravity effects. 

During October 2007, \agile\ detected the source at a flux about a factor of 2
lower than the
September one with no significant variability. Simultaneously, {\textit{Swift}}
observed strong variability (up to a factor $\sim$ 4) in soft X-rays, moderate
variability at optical/UV (less than a factor 2) and approximately constant
hard X-ray flux. Also the different variability observed in optical/UV, soft
and hard X-rays suggests the presence of 2 SSC components in the SED of this object (\cite{Gio}).

\subsection{The Virgo Region: 3C 279 and 3C 273}

Past observations of the Virgo region by the $CGRO$ revealed the presence of two bright and variable
$\gamma$-ray blazars: 3C 273 and 3C 279; therefore, the \agile\, satellite performed dedicated pointings of the
Virgo region for investigating the properties of these two blazars. 

\begin{figure*}[!hhht]
\centering
\sidecaption
\includegraphics[width=12.4cm]{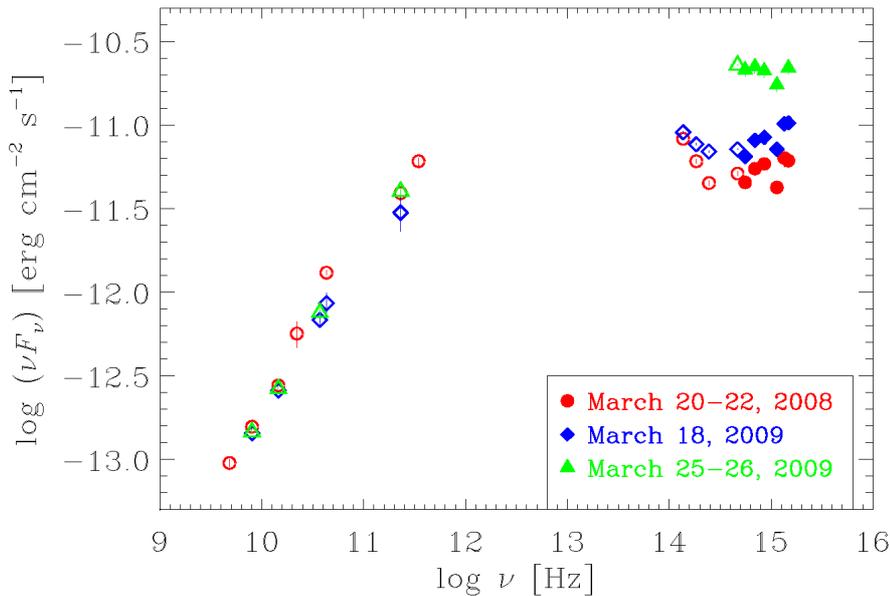}
  \caption{\,SED of the low-energy part of the spectrum of PKS 1510$-$089
  constructed with data collected by GASP-WEBT and {\textit{Swift}}/UVOT
  during March 2008 and March 2009. [Adapted from \cite{DAm09b}].} 
 \end{figure*}

\begin{figure*}[!hhht]
\centering
\sidecaption
\includegraphics[width=12.4cm]{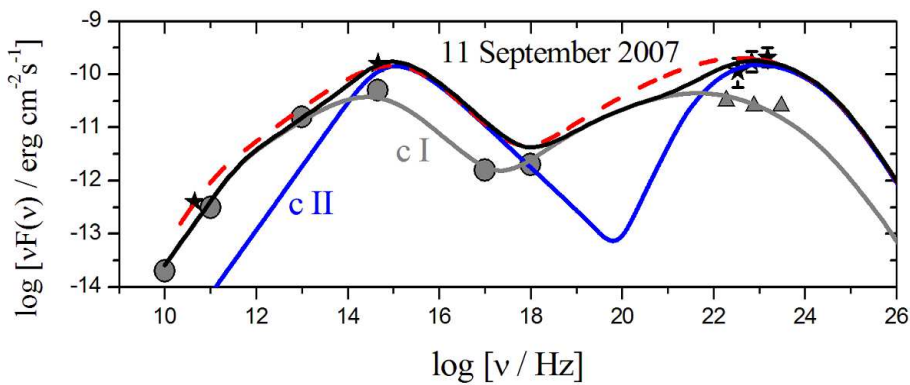}
  \caption{\,SED of S5 0716$+$714 in mid-September 2007 including optical
  GASP-WEBT and $\gamma$-ray \agile\ data (black stars). Historical data
  relative to a ground state and EGRET data are represented with grey
  dots. Curves ``c I'' and ``c II'' represent the two separate
  components. The solid black line and the dashed red lines represent the
  two-components and one-component model, respectively. [Adapted from \cite{Vit09}].} 
 \end{figure*}

3C 279 is the first extragalactic source detected by \agile\ in mid July 2007,
as reported in \cite{Giu09}. The average
$\gamma$-ray flux over 4 days of observation is F$_{\rm E>100\, MeV}$ = (210 $\pm$
38)$\times$10$^{-8}$ ph cm$^{-2}$ s$^{-1}$, a flux level similar to the highest
observed by EGRET and \fermilat\,. The spectrum observed during the flaring
episode by \agile\ is soft with respect to the previous EGRET
observations and this could be an indication of a low accretion state of the
disk occurred some months before the $\gamma$-ray observations, suggesting a
dominant contribution of the external Compton scattering of direct disk (ECD)
radiation compared to the external Compton scattering of the BLR clouds (ECC). As a matter of fact, a strong minimum in the optical band was
detected by REM two months before the \agile\, observations and the reduction
of the activity of the disk should cause the decrease of the photon seed
population produced by the disk and then a deficit of the ECC component with
respect to the ECD, an effect delayed by the light travel time required to the photons to go from the inner disk to the BLR. 

On the other hand, 3C 273 is a very peculiar AGN that shows properties
characteristic of a blazar, like strong radio emission, apparent superluminal
jet motion, large flux variations and SED with the two humps (see \cite{Couv98} for a review), but also other features typical of Seyfert galaxies appear
as well as the broad emission lines, the soft X-ray excess and the big blue
bump. Surprisingly, 3C 273 was discovered to emit in $\gamma$-rays by COS-B in
1976 (\cite{swans}). EGRET pointed this FSRQ several times, not always
detecting it, with an average flux of (15.4 $\pm$ 1.8)$\times$10$^{-8}$ ph cm$^{-2}$ s$^{-1}$ (E $>$ 100 MeV) and only recently
\fermilat\, detected two exceptional $\gamma$-ray outbursts by 3C 273 with peak flux of $\sim$ 1000$\times$10$^{-8}$ ph cm$^{-2}$ s$^{-1}$ (\cite{Abdo}).

\begin{figure*}[hhht!]
\centering
\sidecaption
\includegraphics[width=12.4cm]{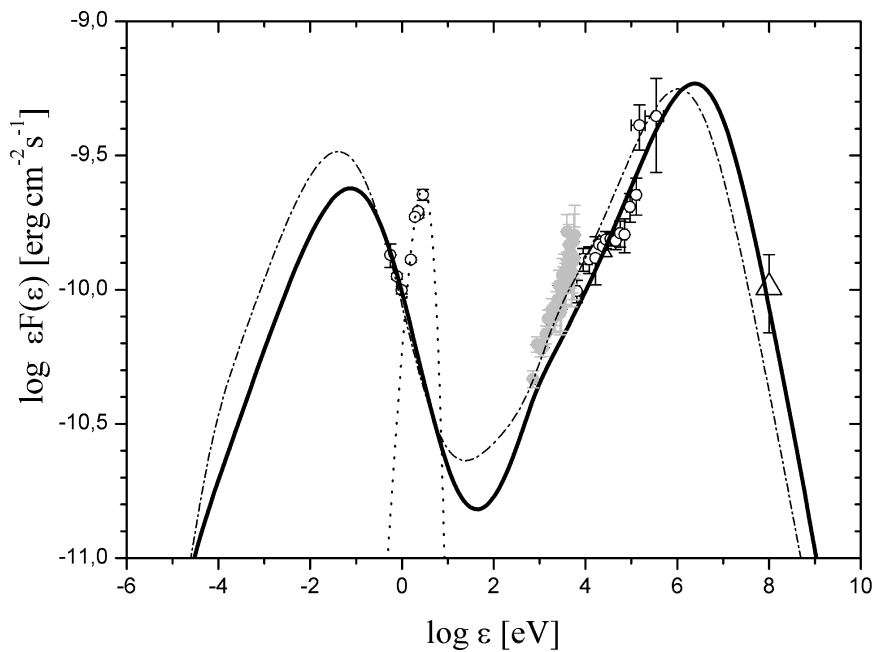}
  \caption{\,SED of 3C 273 for the first (dot-dashed line) and second (solid
  line) week. Triangle is for the AGILE data. The light grey data refers to
  XRT observations, performed in the third week. [Adapted from \cite{Pac09}].} 
 \end{figure*}

We organized a 3-week
multi-frequency campaign between mid-December 2007 and January 2008 on 3C 273
involving REM, RXTE, INTEGRAL, {\it Swift} and \agile\,, with the aim of
studying the correlated variability in the different energy ranges and
time-resolved energy density distribution for each of the 3-weeks from near-IR
to $\gamma$-rays.  During this campaign, whose results are reported in
\cite{Pac09}, the source was detected in a high state in X-rays, with a 5--100
keV flux a factor of $\sim$ 3 higher than the typical value in historical
observations (\cite{Cour03}), whereas the source was detected in $\gamma$-rays
only in the second week, with an average flux of F$_{\rm E>100 \, MeV}$ = (33
$\pm$ 11)$\times$10$^{-8}$ ph cm$^{-2}$ s$^{-1}$. The simultaneous light
curves from near-IR to $\gamma$-rays do not show any strong correlation,
except for an indication of anti-correlated variability between X-rays and
$\gamma$-rays. The SED is well modelled by a leptonic model where the soft
X-ray emission is produced by the combination of SSC and EC models, while the
hard X-ray and $\gamma$-ray emission is due to ECD (Fig. 8). The spectral
variability between the first and the second week is consistent with the
acceleration episode of the electron population responsible for the
synchrotron emission. A possible shift of the IC peak were proposed comparing
the June 1991 campaign with the OSSE observation in September 1994 (\cite{McNaron}). Our multi-frequency observation suggests that this behaviour could be a more general feature of this source, happening on shorter timescales.      

\subsection{TeV blazars: Mrk 421 and W Comae}

\begin{figure*}[!hht]
\sidecaption
\includegraphics[width=11.75cm]{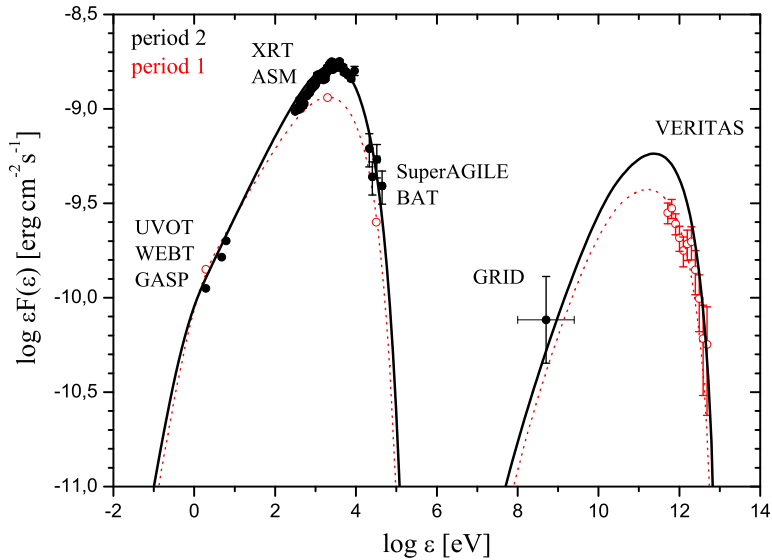}
  \caption{\,SEDs of Mrk 421 obtained by combining GASP-WEBT,
  {\textit{Swift}} (UVOT, XRT, BAT), RXTE/ASM, SuperAGILE, GRID and VERITAS
  data in the period 1 (2008 June 6; empty circles) and period 2 (2008 June
  9--15; filled circles). Both are one-zone SSC
  models. [Adapted from \cite{donnarumma2009}]} 
 \end{figure*}

With the advent of the latest generation of Imaging Atmospheric Cherenkov
Telescopes (IACTs) the number of sources detected in the TeV energy regime has
significantly increased. The majority of TeV sources are galactic, however 28 AGNs are detected until now, but only 8 of the 28 AGN TeV-emitters were detected by
EGRET (\cite{Hart99}) and most of these sources were discovered at TeV
energies only by the new generation of IACTs, therefore the number of TeV blazars
detected contemporaneous at MeV--GeV and TeV energy bands is very low. With the launch of two new $\gamma$-ray satellites, \agile\, and \fermi\,, the
gap in the MeV--GeV domain have been closed giving the possibility to remove the degenerecies in
the modelling of the SEDs of these objects.  
Multi-wavelength campaigns involving $\gamma$-ray \agile\, observations, together with MAGIC and VERITAS
TeV observations of Mrk 421 and W Comae were performed in June 2008.

On 8 June 2008, VERITAS announced the detection of a TeV flare from the IBL
object W Comae (\cite{Sw}), with a three times higher flux with respect to the
flare observed in March
2008 (\cite{Ac08}). About 24 hours later, \agile\ re-pointed towards the source and
detected it with a flux for E $>$ 100 MeV of (90 $\pm$ 32)$\times$10$^{-8}$ ph
cm$^{-2}$ s$^{-1}$, roughly a factor of 1.5 larger than the highest flux
observed by EGRET (see \cite{Ver}). The VERITAS observations triggered a multi-wavelength
campaign including also {\textit{Swift}}, XMM-{\textit{Newton} and GASP observations,
covering the entire electromagnetic spectrum from radio to TeV. The SED
of W Comae during the VHE $\gamma$-ray flare can be modelled by a simple
leptonic SSC model, but the wide separation of the two peaks in the SED requires low ratio of the magnetic
  field to electron energy density ($\epsilon_B = 2.3\times 10^{-3}$), far
  from the equipartition. The SSC+EC model returns magnetic field parameters closer
to equipartition, providing a satisfactory description of
the broadband SED (\cite{Ac09}). 

During the ToO towards W Comae, \agile\ detected also the HBL object Mrk 421, in hard X-rays and $\gamma$-rays. SuperAGILE detected a fast increase of flux
from Mrk 421 up to 40 mCrab in the 15--50 energy band, about a factor of 10 higher
than its typical flux in quiescence (\cite{cos08}), reaching about 55
mCrab on 2008 June 13. This observation was followed by the detection in
$\gamma$-ray by GRID with a flux, F$_{ \rm E>100\, MeV}$ =
(42 $\pm$ 13)$\times$10$^{-8}$ ph cm$^{-2}$ s$^{-1}$, about a factor
3 higher than the average EGRET value, even if consistent with its maximum. An extensive multi-wavelength campaign from optical to TeV energy bands
was organized with the participation of WEBT, $Swift$, RXTE, \agile\,, MAGIC and
VERITAS, as reported in detail in \cite{donnarumma2009}. SuperAGILE, RXTE/ASM and {\it Swift}/BAT show a clear correlated flaring structure
between soft and hard X-rays with a high flux/amplitude variability in hard
X-rays. Hints of the same flaring behaviour is also detected in optical band
by GASP-WEBT. Moreover, \it Swift}/XRT observed the source at the highest 2--10 keV flux
  ever observed, with a peak of the synchrotron at $\sim$3 keV, showing a
  shift with respect to the typical values of 0.5--1 keV. VERITAS and MAGIC
  observed the source on 2008 June 6--8 in a bright
  state at TeV energies, well correlated with the simultaneous peak in X-rays.
The SED can be interpreted
within the framework of the SSC model in terms of a rapid acceleration of
leptons in the jet (Fig. 9). 
An alternative more complex scenario, in the context of the helical jet models, is that optical and X-ray emissions come from different
regions of the jet, with the inner jet region that produces X-rays and is
partially transparent to the optical radiation, whereas the outer region
produces only the low-frequency emission.

\begin{acknowledgements}
This workshop has been supported by the European 
Community Framework Programme 7, Advanced Radio Astronomy in Europe, 
grant agreement no.: 227290.
The AGILE Mission is funded by the ASI with scientific and progammatic partecipation by the
Italian Institute of Astrophysics (INAF) and the Italian Institute of Nuclear
Physics (INFN). FD would like to thank the organizers for the financial support.
\end{acknowledgements}


\begin{thebibliography}{}

\bibitem[Abdo et al. 2010a]{Abdo} Abdo, A. A., et al. 2010a, ApJ, 714, L73

\bibitem[Abdo et al. 2010b]{Abdo2} Abdo, A. A., et al. 2010b, ApJ, 715, 429

\bibitem[Acciari et al. 2008]{Ac08} Acciari, V.~A., et al. 2008 , ApJ, 684, L73

\bibitem[Acciari et al. 2009] {Ac09} Acciari, V.~A., et al. 2009, ApJ, 707, 612  

\bibitem[Blandford $\&$ Rees 1978]{BR} Blandford, R. D.,  $\&$ Rees, M. 1978, in BL Lac
  Objects, ed. A. M. Wolfe, Pittsburgh Press

\bibitem[Blandford $\&$ Payne 1982]{BP} Blandford, R.~D., $\&$ Payne, D.~G. 1982, MNRAS, 199, 883

\bibitem[Bonnoli et al. 2010]{Bon} Bonnoli, G., et al. 2010, MNRAS accepted, [arXiv:1003.3476] 

\bibitem[B\"ottcher 2007]{Bo} B\"ottcher, M. 2007, Ap$\&$SS, 309, 95

\bibitem[Chen et al. (2008)]{Chen} Chen, A. W., D'Ammando, F., et al. 2008, A$\&$A, 489, L37

\bibitem[Costa et al. 2008]{cos08} Costa, E., et~al. 2008, Astronomer's Telegram, 1574

\bibitem[Courvoisier et al. 1998]{Couv98} Courvoisier, T.~J.~L. 1998, A$\&$AR, 9, 1

\bibitem[Courvoisier et al. 2003]{Cour03} Courvoisier, T. J. L., et al. 2003, A$\&$A, 411, L343

\bibitem[D'Ammando et al. 2009]{DAm09a} D'Ammando, F., et al. 2009, A$\&$A, 508, 181

\bibitem[D'Ammando et al. 2010]{DAm09b} D'Ammando, F., et al. 2010, to be submitted to A$\&$A

\bibitem[Donnarumma et al. (2009a)]{donnarumma2009} Donnarumma, I., et~al. 2009a, ApJ, 691, L13

\bibitem[Donnarumma et al. (2009b)]{Don09} Donnarumma, I., et al. 2009b, ApJ, 707, 1115

\bibitem[Giommi et al. 2006]{Giommi2006:3C454_Swift}
{Giommi}, P. et~al. 2006, A$\&$A, 456, 911

\bibitem[Giommi et al. 2008]{Gio} Giommi, P., et al. 2008, A$\&$A, 487, L49

\bibitem[Giuliani et al. (2009)]{Giu09} Giuliani, A., D'Ammando, F., et al. 2009, A$\&$A, 494, 509

\bibitem[Hartman et al. 1999] {Hart99} Hartman, R. C., et al. 1999, ApJS, 123, 79

\bibitem[Hartman et al. 2001] {Har01} Hartman, R. C., et al. 2001, ApJ, 553, 683

\bibitem[Kataoka et al. (2008)]{Ka} Kataoka, J., et al. 2008, ApJ, 672, 787

\bibitem[McNaron-Brown et al. 1997]{McNaron} McNaron-Brown, K., et al. 1997, ApJ, 474, L85

\bibitem[Meier 2002]{meier2002} Meier, D.~L. 2002, NewAR, 46, 247

\bibitem[Nilsson et al. (2008)]{Ni} Nilsson, K., et al. 2008, A$\&$A, 487, L29 

\bibitem[Pacciani et al. (2009)]{Pac09} Pacciani, L., et~al. 2009, A$\&$A, 494, 49

\bibitem[Pacciani et al. 2010]{Pac10} Pacciani, L., et al. 2010, ApJ, 716, L170

\bibitem[Pian et al. 2006]{Pian2006:3C454_Integral} {Pian}, E., et~al. 2006, A$\&$A, 449, L21 

\bibitem[Pucella et al. 2008]{Puc08} Pucella, G., et al. 2008, A$\&$A, 491, L21

\bibitem[Remillard 2005]{remillard} Remillard, R. 2005, Astronomer's Telegram, 484

\bibitem[Striani et al. 2010]{striani2010} Striani, E., et al. 2010, ApJ accepted, [arXiv:1005.4891]

\bibitem[Swanenburg  et al. 1978]{swans} Swanenburg, B.~N., et~al. 1978, Nature, 275, 298

\bibitem[Swordy 2008] {Sw} Swordy, S. 2008, Astronomer's Telegram, 1565

\bibitem[Tavani et al. 2009] {Tav09} Tavani, M., et al. 2009, A$\&$A, 502, 995

\bibitem[Ulrich et al. 1997]{Ul} Ulrich, M., et al. 1997,  ARA$\&$A, 35, 445

\bibitem[Urry $\&$ Padovani 1995]{UP} Urry, C. M., $\&$ Padovani, P. 1995, PASP, 107, 803

\bibitem[Vercellone et al. 2008]{Vercellone2008} {Vercellone}, S., et~al. 2008, ApJ, 676,
    L13

\bibitem[Vercellone et al. (2009)]{Ver09} Vercellone, S., et al. 2009, ApJ, 690, 1018

\bibitem[Vercellone et al. 2010]{V18} Vercellone, S., D'Ammando, F., et al. 2010, ApJ, 712, 405

\bibitem[Verrecchia et al. 2008]{Ver} Verrecchia, F., et al. 2008, Astronomer's Telegram, 1582

\bibitem[Villata et~al. 2006]{vil06}
{Villata}, M., {Raiteri}, C.~M., et~al. 2006, A$\&$A, 453, 817

\bibitem[Vittorini et al. 2009]{Vit09} Vittorini, V., et al. 2009, ApJ, 706, L1433

\end{thebibliography}
\end{document}